# Numerical Study on the effect of port geometry of intake manifold in a Steam Wankel Expander

Auronil Mukherjee[1], Satyanarayanan Seshadri[1]*

[1]Energy and Emissions Research Group, Indian Institute of Technology Madras, Department of Applied Mechanics,
Chennai-600036, Tamil Nadu, India
Tel: +91 44 2257 4078, e-mail: satya@iitm.ac.in

* Corresponding Author

**ABSTRACT**

*A Wankel steam expander has numerous advantages over other positive displacement machines as an expansion device due to its high power-to-weight ratio, compactness, lower noise, vibration, and potentially lower specific cost. In these expanders, the pressure drop of steam during admission through rotary valves is inevitable across the intake manifold of the expander during admission duration. These pressure losses during intake change the design pressure ratio across the expander, reducing power output by a reasonable margin of 20 to 30%. Therefore, reducing it to improve the net power output is crucial. The goal of the present research is twofold. In the first part, we estimate the pressure losses across the intake manifold of the expander for a rectangular port geometry. In the second part, a trapezoidal port profile of the same hydraulic diameter is designed for the intake manifold to reduce the intake losses, thereby delivering a higher power output. The thermodynamic analysis, with state points from REFPROP, is performed for the theoretical pressure-volume cycle of the expander to derive the boundary conditions for a CFD model. We observe that the trapezoidal port significantly reduces the pressure losses by around 50%, delivering around 7 to 21% higher net power output. The increase in isentropic efficiency is about 14% over a range of rotational speeds from 1200 to 3000 RPM. Further, we study the effect of different fluid flow and turbulent parameters on the expander's pressure loss, power output and isentropic efficiency.*

1. INTRODUCTION

Renewable energy, which includes sun, biomass, low temperature geothermal and waste heat recovery, is a solution to the problem of climate change by utilising various energy heat sources. All these are examples of low-grade heat technologies for generating useful power.

The Organic Rankine Cycle is a suitable technology for utilising these sources [1-3]. When it comes to small plants, one of the essential components of an ORC cycle is the expander, whose performance and costs result in a better plant efficiency than turbines[3]. The output power of these systems varies from a few kW to several MW, requiring the selection of an appropriate expander for each power output. Scroll expanders are mainly utilised for home applications (1–10 kW) [4-8]. On the other hand, high efficiency can only be achieved in bigger power plants by using turbines, which suffer a significant drop in efficiency when operating outside their design parameters. Reciprocal volumetric expanders are a good technique in the mid-range of power output (10–100 kW) [9]. Volumetric expanders, among other benefits, can be used with nearly any fluid, regardless of the moisture content at the end of the expansion. Dumont et al.[10] tested small-scale expanders for application in an ORC system using R245fa as the working fluid. They reported an effective isentropic efficiency of 81 per cent for a fixed-speed scroll. They observed low efficiency of 53% for screw and reciprocating expanders. For full-load Organic Rankine cycle applications, Zywica et al.[11] suggested using turbines, screw, and scroll instead of reciprocating and Wankel expanders. They also advocate a radial inflow turbine in the organic Rankine cycle (ORC) for low-flow and high-pressure differential scenarios due to its excellent isentropic efficiency. Gopal and Seshadri[12] stated that impulse turbines might not be the best choice for expanding steam produced by saturated steam boilers in process applications.

Glovanni et al.[13] A detailed examination of a Wankel engine as an expansion device with air as the working fluid in a US naval experiment undertaken in the 1970s. They claimed that mechanical efficiency is 70% and energy efficiency is 30% to 60%. Several investigators[14-16] have demonstrated that a rotary expander, designed from a modified Wankel engine, would be suitable for this usage because of its compactness, economy and capability of operating with different fluids in various operating conditions. Furthermore, when the Wankel engine is changed to an expander, the typical drawbacks of the Wankel engine, namely high specific consumption and pollutants, are nearly eliminated [16].

Badr et al.[14] examined the Wankel Expander for power generation using the Rankine steam power cycle. For the commercially available Wankel engines from Mazda and Curtiss-Wright, they created a numerical model and described the performance of the expansion devices for varied boiler pressures. Two input and two exhaust ports were chosen for the design, resulting in two power pulses every revolution. Most studies used the Wankel engine as a combustion engine using various working fluids such as gasoline, methane, octane, hydrogen, diesel, and petrol [17]. Many others used it as a compressor or expander device [18-19]. Antonelli et al. [20] compared simulation and experiments regarding supplied torque, working fluid mass flow rate, and indicated pressure for an Organic Rankine Cycle. The results were validated using compressed air. They used a lumped parameter numerical model to consider the losses due to leakage, friction, and heat transfer. Using pentane as the working fluid, they achieved an isentropic efficiency of around 85% and a thermal cycle efficiency of 10% at 80 °C. As an alternative to ports, Francesconi et al.[21-26] used valves in the Wankel expander, with intake and exhaust valves running at a 1:2 timing ratio, which is the rotational speed ratio of the valves and crankshaft. Investigators in [27] tested and published the isentropic efficiency of the Wankel expander for several organic fluids at various cut-offs, with a peak isentropic efficiency of the order of 88 % for R600a. A CFD simulation of a Wankel expander with compressed air as the working fluid is presented by Sadiq et al. [28]. For a two-stage expansion, the highest isentropic efficiency was calculated to be 91%. Simulated inlet pressures of 4 and 6 bars were used in the investigation. It was claimed that the isentropic efficiency for single-stage expansion from 4 bar was lower than 6 bar. Later, however, it was revealed that the two-stage expansion from 6 bar had higher power output and isentropic efficiency than both single-stage expansion scenarios.

Most of the previous research has concentrated on the performance parameters of Wankel expander in various operating conditions, operating it with multiple working fluids, and modelling leakage and thermal losses. Observing the nature of flow dynamics and discharge coefficients through intake and exit valves has also been the subject of several studies. Researchers have stated friction, leakage, initial condensation, and heat loss as the contributing parameters affecting the expander's net power output and efficiency [12]. However, there was no extensive analysis of net power output losses due to pressure losses across the intake manifold during admission through rotary valves and the effect of intake port shape on pressure loss and net power output recovery. These considerations make it worthwhile to evaluate the amount of the losses across the intake manifold, the effect of intake port geometry and their impact on the expander's net power production over a wide range of RPMs, as done in the current study for a Wankel expander prototype established in the Energy and Emissions Research Group (EnERG) of IIT Madras [12].

When saturated steam exits the intake valve port, it encounters a sudden expansion near the valve exit. Following that, it travels a finite distance to enter the expander chamber via the port in the rotor housing. In the following sections, a thermodynamic state point analysis of the expander is performed in Python 3.8 coupled with the NIST-REFPROP® database to visualise the pressure-volume variation inside the expander chamber in agreement with reported literature [12]. Subsequently, a thorough numerical investigation is performed to estimate the pressure loss variation across the intake manifold with the rotor angle at five different RPMs ranging from 1200 to 3000 for an existing rectangular intake port. The power losses are calculated for the ideal power output of the expander. Leakage and thermal losses are neglected, and the steam is assumed to be dry saturated before the beginning of the expansion process. The numerical solver is validated using established pressure drop correlations and expressions reported in the literature by Idelchik [29] and Duan *et al*. [30]. A trapezoidal port is designed of an equivalent area to the rectangular port to recover the associated pressure losses, delivering higher net power output and isentropic efficiency in the range of RPMs.

The novelty of the work is to highlight the impact of the port profiles on intake pressure losses and the net power production of the expander. The effect and variation of various fluid flow parameters such as skin friction coefficient, wall shear stress, and turbulence kinetic energy are also investigated for both port profiles to provide a detailed understanding of the role of these parameters during steam intake of a Wankel expander.

## 2. WANKEL EXPANDER GEOMETRIC DETAILS

The Wankel expander prototype comprises a static housing, a triangular rotor, an eccentric output shaft, and two rotating intake exhaust valves. The shape of the rotor housing and flanks is determined by the rotor radius R and the eccentricity e of the output shaft. A schematic of the Wankel geometry is shown in Fig. 1. The rotor rotates around its centre and moves along the eccentric shaft radius, e. The shaft completes three revolutions around the eccentric circle while the rotor rotates one revolution around its centre. The design parameters of the Wankel expander are shown in Table 1.

The parametric equations for housing are as follows:

$$x_h = e\cos 3\theta + r\cos\theta \tag{1}$$

$$y_h = e\sin 3\theta + r\sin\theta \tag{2}$$

The following are the equations for the rotor's shape:

$$x_r = r\cos 2\upsilon + \frac{3e^2}{2R}(\cos 8\upsilon - \cos 4\upsilon) \pm e\left(1 - \frac{9e^2}{R^2}\sin^2 3\upsilon\right)^{\frac{1}{2}}(\cos 5\upsilon + \cos\upsilon) \tag{3}$$

$$y_r = r\sin 2\upsilon + \frac{3e^2}{2R}(\sin 8\upsilon - \sin 4\upsilon) \pm e\left(1 - \frac{9e^2}{R^2}\sin^2 3\upsilon\right)^{\frac{1}{2}}(\cos 5\upsilon + \cos\upsilon) \tag{4}$$

The intervals of $\upsilon$ in expressions (3) and (4) are as follows:

$$\upsilon = [\tfrac{\pi}{2}, \tfrac{5\pi}{6}], [\tfrac{11\pi}{6}, \tfrac{13\pi}{6}], [\tfrac{19\pi}{6}, \tfrac{21\pi}{6}]$$

**Table 1**: Design Parameters for the Wankel Expander

| Parameter | Value | Unit |
|---|---|---|
| R | 80.0 | mm |
| e | 10.5 | mm |
| $b$ | 71.0 | mm |

## 3. THERMODYNAMIC ANALYSIS

The Wankel Expander is built for a design pressure ratio. The boiler and condenser pressures of a Rankine Cycle system, in which the Wankel must function as an expansion device, are considered while determining the intake and exhaust pressure limits. The expander's pressure ratio is $r_p$. $P_b$ and $P_c$ are the boiler and condenser pressures, respectively.

$$r_p = \frac{P_b}{P_c} \qquad (5)$$

Figure 2 depicts a schematic of a theoretical pressure-volume variation for the expander, which repeats twice for each rotor rotation, resulting in six power strokes per full rotation [14]. The valve and shaft speed ratio is 1:1[12]. Table 2 shows the expansion device's operational parameters.

When the rotor is at the clearance volume, the cylindrical intake rotary valve port remains entirely closed. The port begins to open when the steam admission process begins, progresses to a fully open stage, and then closes completely at the cut-off. During admission, the steam is in a saturated vapour state, and during expansion, it is in a liquid-vapour mixture state. Equating the entropy at the beginning and end of expansion yields the steam quality at the maximum chamber capacity. The NIST-REFPROP® database calculates the specific volume at the end of the expansion. The maximum and minimum volume of the Wankel chamber can be calculated by taking the extremum of the expression (6). The cut-off, compression ratio and isentropic index are calculated using the adiabatic law at the expansion process's beginning and end using a similar approach as demonstrated in [12]. The mass of steam expanded in the chamber is calculated using mass conservation, yielding the mass of steam intake in a single cycle. An ideal pressure-volume variation in a single working chamber of the expander is schematically depicted in Figure 2.

**Table 2:** Operating Parameters of the Wankel Expander

| Parameter | Range/Value | Unit |
|---|---|---|
| $P_b$ | 10 | bar |
| $T_{in}$ | 453 | K |
| $P_c$ | 3 | bar |
| $N$ | 1200-3000 | RPM |

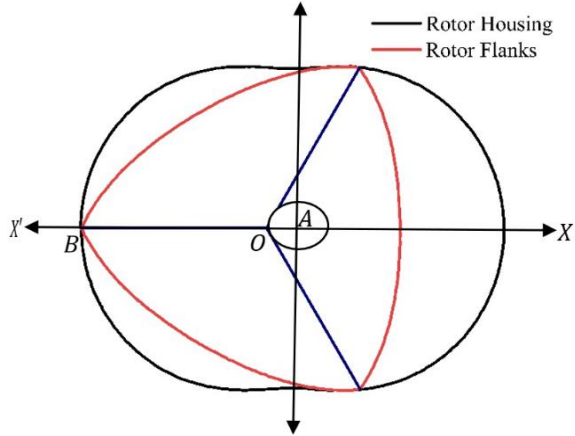 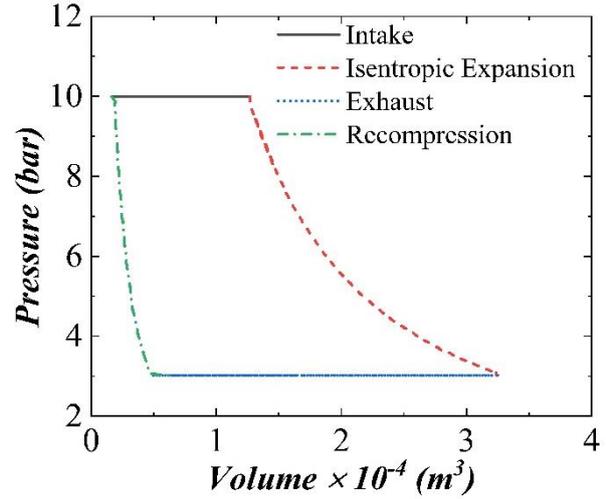

**Figure 1:** Schematic of Wankel Geometry.  **Figure 2:** Ideal pressure volume variation in a single chamber.

### 3.1 VOLUME AND MASS FLOW RATE MEASUREMENT

During operation, the triangle rotor inside the housing rotates, causing a continual change in the volume of the working chamber. The rotor housing width is multiplied by the side area contained by a rotor contour side and the inner surface of the epitrochoid to get this volume. The expander design parameters and the shaft angle determine the working chamber volume. It is stated as follows [31]:

$$V(\theta) = \frac{\pi}{3}e^2 + eR\left(2\left(1 - \frac{9e^2}{R^2}\right)^{\frac{1}{2}} - \frac{3\sqrt{3}}{2}\sin\left(\frac{2\theta}{3} + \frac{\pi}{2}\right)\right) + \left(\frac{2}{9}R^2 + 4e^2\right)\sin^{-1}\left(\frac{3e}{R}\right) \quad (6)$$

Using the derivative of expression (6) and multiplying with the shaft speed yields the volume flow rate of steam in the chamber. The product of the volume flow rate and the density of the admitted steam yields the corresponding mass flow rate during admission till cut-off. The following is the mathematical approach:

$$\frac{dV}{dt} = \frac{\partial V}{\partial \theta} * \frac{\partial \theta}{\partial t} = \omega \frac{\partial V}{\partial \theta} \quad (7)$$

$$\frac{dm}{dt} = \rho_{ad}\omega \frac{\partial V}{\partial \theta} \quad (8)$$

Figure 3(a-b) illustrates the comparative variation of the volume of the working chamber and its corresponding derivative with the rotor angle. The present study's results and cited literature seem to match reasonably.

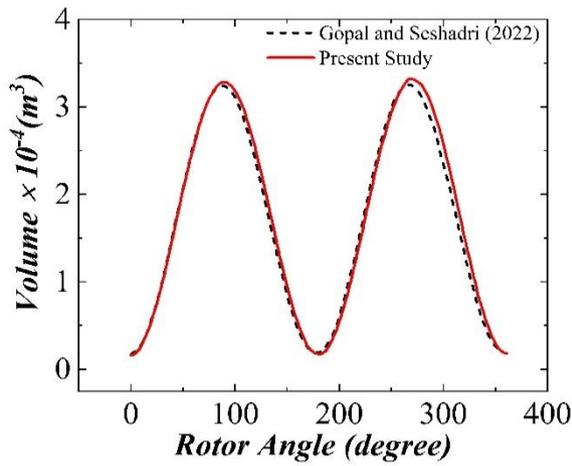
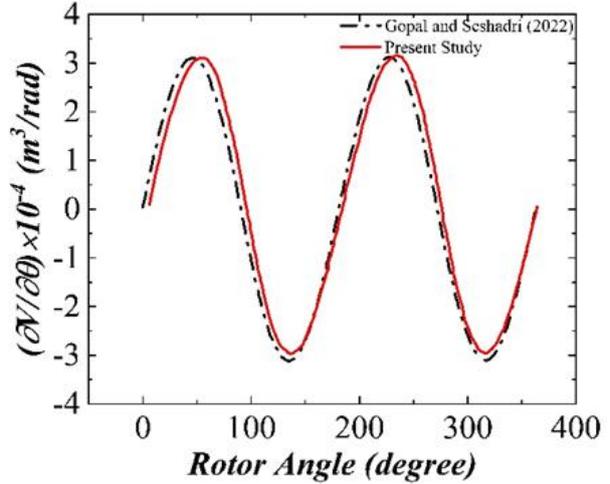

**Figure 3(a):** Comparison of volume-rotor angle variation of expander chamber volume with published literature.

**Figure 3(b):** Comparison of the corresponding derivative –rotor angle with published literature.

Based on the thermodynamic calculations, the volume of the working chamber at the steam cut-off for the design pressure ratio is calculated. The variation of mass flow rate with rotor angle at various rotational speeds is investigated during the admission duration. The fluctuation of the mass flow rate inside the chamber volume is depicted in Figure 4. The mass flow rate shows an increasing tendency from the beginning of the intake to the end due to the sinusoidal structure of the volume-theta expression (6). Before the intake valve port closes, the flow rate reaches a maximum, and the magnitude of the flow rate is directly proportional to the shaft speed.

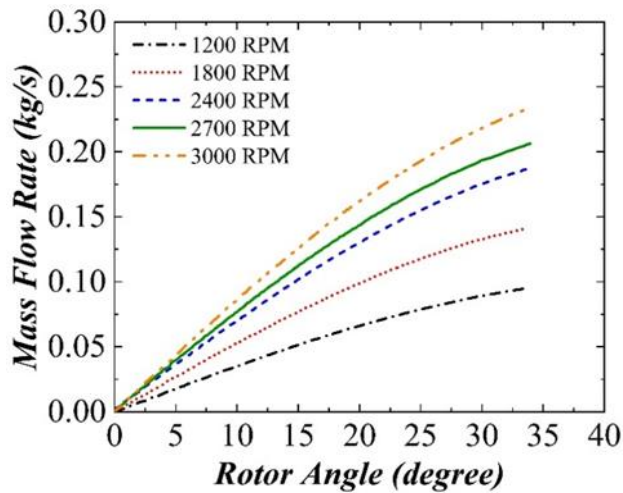

**Figure 4:** Variation of Mass Flow Rate with rotational speed.

### 3.2 INTAKE MANIFOLD AND VALVE PORT GEOMETRY DETAILS

The intake system of the expander consists of a pair of rotary valves rotating in the gear ratio, with ports on them. The saturated steam from the boiler enters the expander chamber through the port of the rotary intake valves. On exit from the valve port, the steam expands and travels a finite distance to enter the chamber volume through the intake port in the rotor housing. The valve port remains fully closed at the expander clearance volume. The valve shaft starts to rotate at the onset of the admission process, which marks the opening of the valve port. It reaches a fully opened state and finally closes at the steam cut-off point. The intake mass flow rate of the steam is minimum at the beginning of the opening of the port, and it reaches a maximum value just before the cut-off. The port on the housing side always remains open. A photograph of the existing valve port prototype of the Wankel expander is illustrated in Figure 5(a).

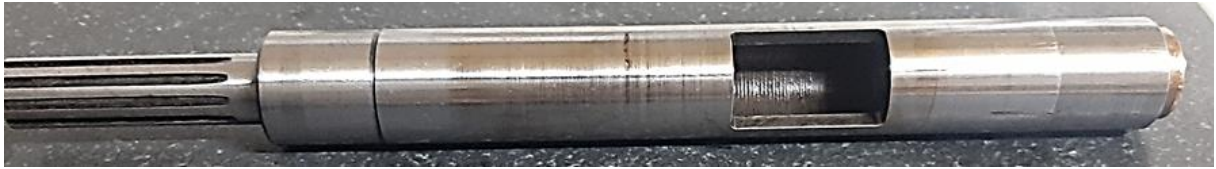

**Figure 5(a):** Photograph of the intake valve- shaft with rectangular port.

The intake port has a rectangular geometry. A two-dimensional schematic of the rectangular port is shown in Figure 5(b). The port starts to open along its width during the admission process.

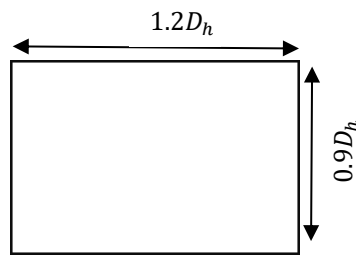

**Figure 5(b):** Schematic of the rectangular valve port geometry.

Design of the width of the port is done keeping in mind the cut-off point of the specified pressure ratio of the Wankel Expander.

The investigations of subsequent sections of the paper are threefold. The first part provided a detailed numerical analysis of the pressure losses across the intake manifold for the above port geometry and its effect on the net power output of the expander using ANSYS Fluent 19.2®. A grid independence test is performed, and the results are validated against existing literature. In the second part, we develop a novel trapezoid port of equivalent hydraulic diameter to the initial rectangular port and investigate the associated pressure losses and their effect on the net power output compared to the previous rectangular port. Finally, the third part presents a detailed comparative analysis of the net power output and different fluid flow and turbulence parameters during the entire duration of opening and closing for both the port profile.

### 3.3 COMPUTATIONAL FLUID DYNAMICS STUDY

#### 3.3.1 Three-dimensional computational details for rectangular port

The computational domain consists of the intake manifold of the Wankel Expander, which extends from the intake valve port's departure to the port in the rotor housing. 3D models are created in ANSYS Design modular for each angular instant of the rotor from admittance to cut-off. The domain's inlet area is modified according to the rotor angle. At a rotor angle of 9º, Figure 6(a) shows a schematic of the flow domain. The red arrow indicates the flow stream's direction. The intake rotary valve rotates during admission, causing the port to open and close along its width. The domain is discretised using a hexahedral mesh to be locally organised and retain orthogonal grids along the wall's normal direction.

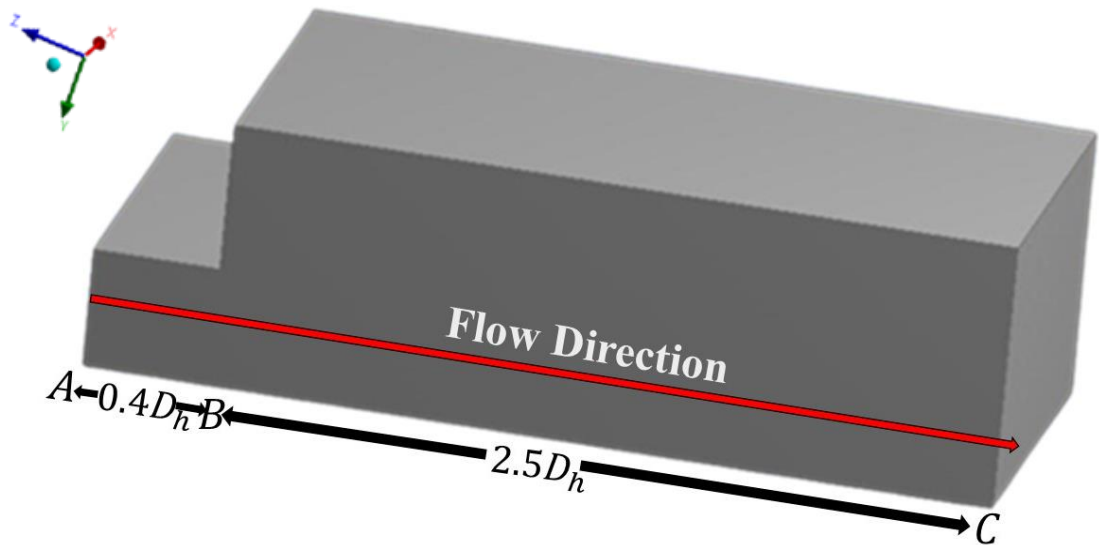

**Figure 6(a):** Schematic of the computational domain with rectangular port.

Figure 6(b) depicts a schematic of the mesh geometry.

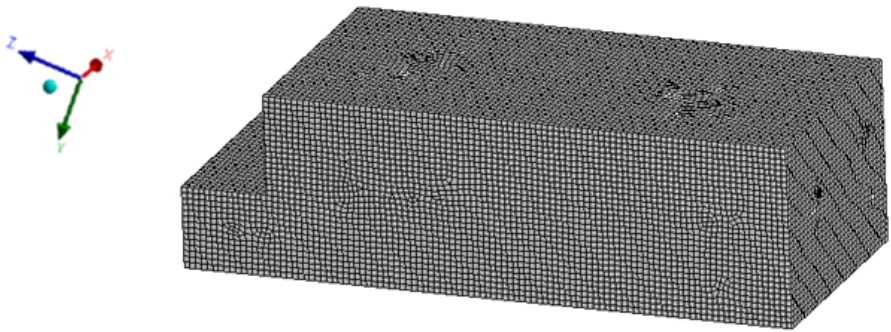

**Figure 6(b):** Schematic of the meshed geometry.

A two-dimensional schematic of the proposed trapezoidal port is illustrated in Figure 7(a). The port profile is designed for an equivalent hydraulic diameter of the rectangular port and starts to open from the shorter side at the beginning of the intake duration. In this work, we take the square root of the cross-sectional area as the hydraulic diameter in agreement with the reported literature by Duan et al.[30].

### 3.3.2 Three-dimensional computational details for trapezoidal port

A similar modelling and meshing methodology are carried out for the intake manifold with a trapezoidal port profile. Figure 7(b) depicts a schematic of the computational domain for the trapezoidal port with a rotor angle of 9º

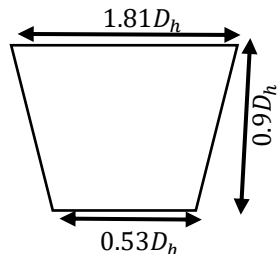

**Figure 7(a):** Schematic of the trapezoid port geometry.

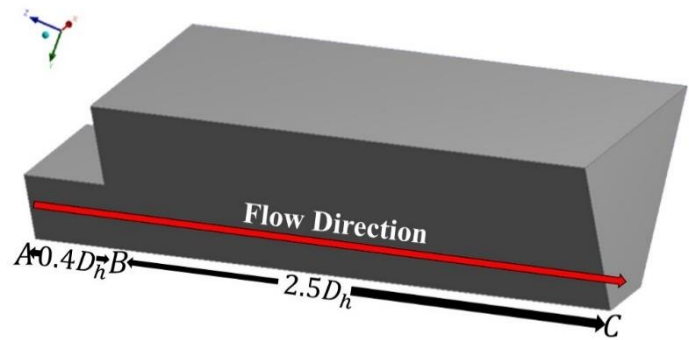

**Figure 7(b):** Schematic of the computational domain with trapezoidal port.

### 3.3.3 Numerical Method and Governing Equation

The governing equations for fluid flow modelling based on mass, momentum, and energy conservation are given in the vector form in equations (9), (10) and (11). The flow is incompressible with no slip on the wall boundary and an adiabatic condition. The atmospheric pressure and ambient temperature are set to 1.013 bar and 300K, respectively. The transient pressure-based solver is chosen at each rotor angular moment to ensure a time-dependent flow solution.

The mass flow rate continuously fluctuates during the admission duration, as illustrated in Figure 3, which results in continuous fluctuation of the mean inlet velocity during the intake duration. Eddies superimposed on the laminar flow characterise turbulence.

### 3.3.4 Choice of relevant turbulence model

The Reynolds Averaged Navier Stokes (RANS) model is used in the present investigation due to its simplicity, robustness, and wide range of applications. However, choosing an appropriate RANS model is necessary to model the associated turbulence and eddies. The Spalart-Allmaras model performs poorly for 3D flow, involving strong separation and free shear flows, although it is economical for large meshes. The most used standard $k$-$\varepsilon$ model has its limitations. It performs poorly for complex flows involving adverse pressure gradient and separation and modelling strong streamlined curvatures. The standard $k$-$\omega$ model provides better performance for free shear flow, wall-bounded boundary layer and flows under an unfavourable pressure gradient. However, it is more suited for low Reynolds number flows. The RNG and realisable $k$-$\varepsilon$ models are highly suitable for complex shear flows, which involve boundary layer separation, rapid strain, vortices, and locally transitional flows. Based on the underlying flow physics of the present problem and the geometry of the computational domain, this model is used to simulate turbulence, which is also in agreement with the reported literature [28]. The renormalisation group $k$-$\varepsilon$ (RNG) model is used to simulate the turbulence, and numerical simulation is done using the 'coupled' pressure-velocity approach. The residuals of continuity, momentum, $k$ and $\varepsilon$ are of the order of 10e-4, respectively.

The three-dimensional Continuity, Momentum and Energy equations for incompressible flow with constant fluid properties are as follows:

$$\nabla \cdot \vec{U} = 0 \tag{9}$$

$$\rho \frac{D\vec{U}}{Dt} = \rho \vec{g} - \nabla P + \mu \nabla^2 \vec{U} \tag{10}$$

$$\rho c_p \frac{DT}{Dt} = \rho \dot{q} + k\nabla^2 T + \varphi \tag{11}$$

The transport equations for the RNG $k$-$\varepsilon$ Model are:

$$\frac{\partial}{\partial t}(\rho k) + \frac{\partial}{\partial x_i}(\rho k u_i) = \frac{\partial}{\partial x_j}\left(\alpha_k \mu_{eff} \frac{\partial k}{\partial x_j}\right) + G_k + G_b + \rho \varepsilon - Y_M + S_k \tag{12}$$

$$\frac{\partial}{\partial t}(\rho \varepsilon) + \frac{\partial}{\partial x_i}(\rho \varepsilon u_i) = \frac{\partial}{\partial x_j}\left(\alpha_\varepsilon \mu_{eff} \frac{\partial \epsilon}{\partial x_j}\right) + C_{1\varepsilon} \frac{\varepsilon}{k}(G_k + C_{3\varepsilon} G_b) - C_{2\varepsilon} \rho \frac{\epsilon^2}{k} - R_\varepsilon + S_\varepsilon \tag{13}$$

The generation of turbulence kinetic energy due to mean velocity gradients is denoted by $G_k$ in the preceding two equations, and the generation of turbulence kinetic energy due to buoyancy is given by $G_b$. The contribution of variable dilatation in compressible turbulence to the overall dissipation rate is represented by $Y_M$. The inverse effective Prandtl numbers for $k$ and $\varepsilon$ are $\alpha_k$ and $\alpha_\varepsilon$, respectively. $S_k$ and $S_\varepsilon$ represent the user-defined source terms.

### 3.3.5 Initial and Boundary Conditions

The inlet of the intake system of the expander is connected to a high-pressure supply boiler, supplying saturated steam at a gauge pressure of 10 bar. A constant pressure boundary condition is incorporated in the inlet of the intake manifold. The corresponding boundary conditions at the inlet are as follows:

$$P = P_b, \quad T = T_{in} \tag{14}$$

A mass flow rate boundary condition is applied at the flow domain's outlet. Using equations (6), (7), and (8), the magnitude of the mass flow rate, which is a function of the expander's shaft angle, is adjusted depending on the inlet area of the flow domain. The upper and lower walls of the manifold are subjected to no-slip and adiabatic boundary conditions. The manifold outlet, towards the chamber volume side, always remains open. A variation of the mean inlet velocity with the rotor angle is illustrated in Figure 8(a-b) for three different RPMs for the rectangular and trapezoidal ports. The plot shows that the velocity remains constant during the opening for the rectangular port, while it decreases slowly for the trapezoid port. The velocity magnitude follows a rapid increase during both ports' closing duration.

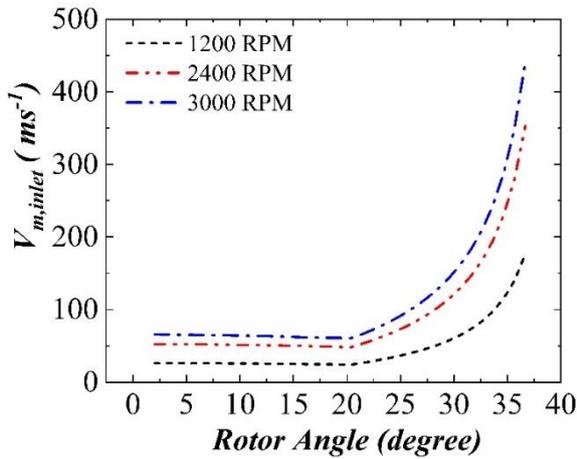 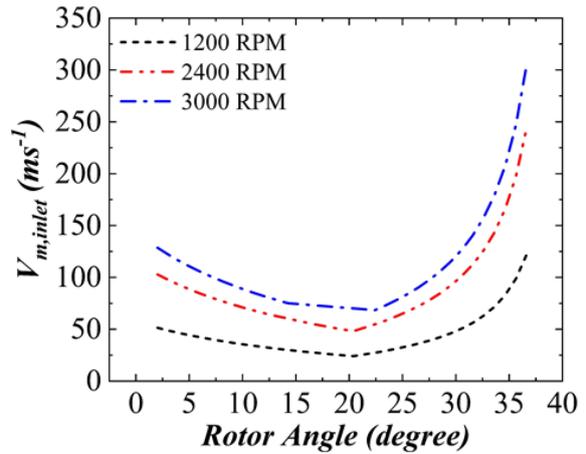

**Figure 8(a):** Variation of mean inlet velocity of the flow domain with rotor angle for the rectangular port.

**Figure 8(b):** Variation of mean inlet velocity of the flow domain with rotor angle for the trapezoidal port.

**Table 3: Details of the boundary conditions of the computational domain**

| Location | Type of Boundary | In mathematical form | Magnitude | Unit |
|---|---|---|---|---|
| Inlet | Pressure | $P_b = $ constant | 10 | *bar* |
| Outlet | Mass Flow Rate | $\dot{m} = f(\theta, \omega)$ | 0.00475-0.16 | *kg/sec* |
| Walls | No slip | $U = 0$ | - | - |

### 3.3.6 Mesh Sensitivity Analysis

A grid sensitivity analysis establishes the best mesh size for numerical simulations. The pressure loss at a rotor angle of 36 degrees just before the cut-off, at a rotational speed of 1200 RPM, is calculated for five different element sizes. The error is calculated based on the most refined grid, and the residual is found to be less than 10e-03. As a result, for all other angular instants and shaft speeds, this mesh size is proven to be optimal. Figure 9 illustrates the variation of pressure losses at five different mesh sizes.

### 3.3.7 Model Validation

The expander's intake manifold extends from the rectangular-shaped port on the intake valve to the port on the rotor housing. We used established correlation and mathematical expression [29-30] to validate the numerical results and the governing flow physics. The net head loss is calculated from the inlet to the outlet of the flow domains at various rotor angles and determining the corresponding pressure drop. Saturated steam entering the intake manifold experiences frictional head losses and loss due to abrupt expansion at point B in Figure 6(a). We estimate the head loss in regions AB and BC of the computational domain using the modified Colebrook-White equation (15) proposed by Duan et al.[30].

$$fRe_{\sqrt{A}} = \left(3.6 \log(0.2047 \left(\frac{\Delta}{\sqrt{A}}\right)^{\frac{10}{9}} + \frac{6.115}{Re_{\sqrt{A}}})\right)^{-2} Re_{\sqrt{A}}$$

(15)

The velocity distribution throughout the length of the manifold upstream of an abrupt expansion at point B, is never uniform under practical conditions. We utilise the expression proposed by Idelchik [29], to compute the local resistance coefficient for a high-speed flow with a non-uniform velocity distribution at a large *Re*. The magnitude of the losses at different rotor angular instants is calculated for a rotational speed of 2400 RPM. Figure 10 depicts a comparison plot of pressure loss as a function of rotor angle using both approaches. It is observed that numerical results follow similar trends and reasonably match those obtained using the analytical approach.

### 3.3.8 Error Analysis

Root mean square analysis (RMSE) is performed to compute the deviation between the numerical and the correlation results using the following expression (16)

$$RMSE = \sqrt{\sum_{i=1}^{n} \frac{(\hat{y}_i - y)^2}{n}} \qquad (16)$$

Here $\hat{y}_i$ are the values obtained through numerical simulation, and $y$ are the values obtained through the correlations and expressions established in the literature[29-30]. The total number of data points is denoted by $n$. Both the results follow similar trends and have a good match with a root mean squared error magnitude of 5%.

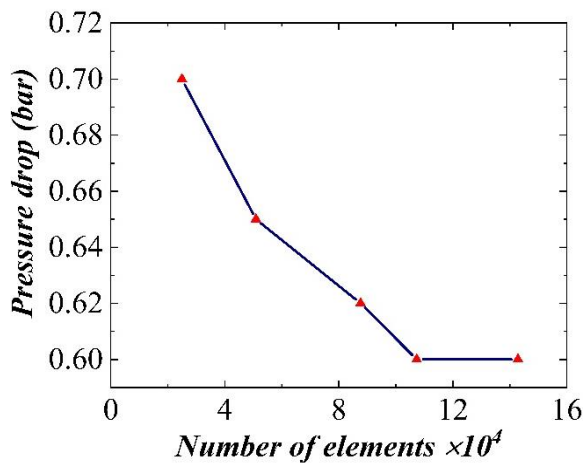 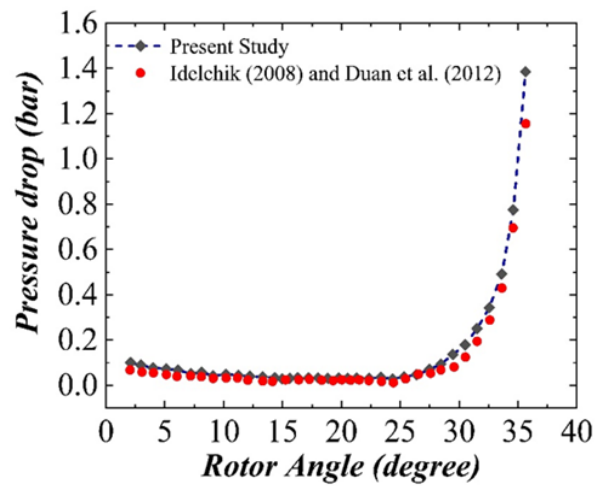

**Figure 9:** Schematic of the grid independence test for different mesh sizes

**Figure 10:** Numerical variation of pressure loss during intake compared to literature

### 4. RESULTS AND DISCUSSIONS

### 4.1 Rectangular Port Results

Pressure loss across the intake manifold for the rectangular port profile during admission duration for a range of rotational speeds ranging from 1200 to 3000 RPM are presented here. The port starts to open along its width at the expander clearance volume.

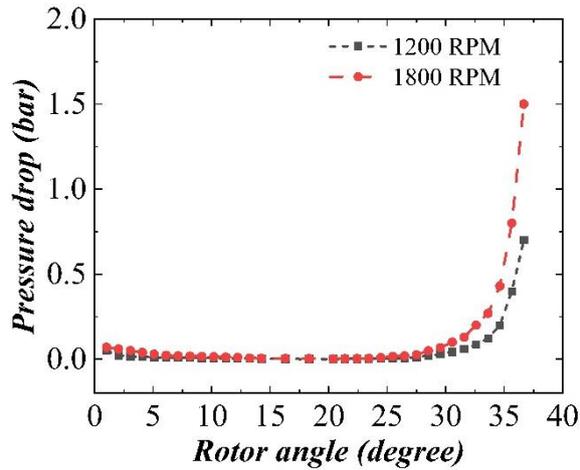 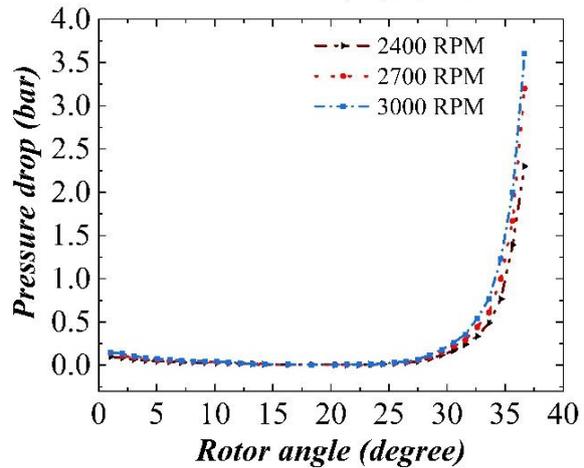

**Figure 11(a):** Pressure drop variation with rotor angle at 1200 and 1800 RPM.

**Figure 11(b):** Pressure drop variation with rotor angle at 2400, 2700, and 3000 RPM.

The pressure drop follows a decreasing trend as illustrated in Figure 11(a-b) during the valve port's opening duration, which ranges from 0 to 15 degrees of rotor angle. The port reaches a fully opened position roughly midway during the intake process. The pressure losses reach a minimum value at this instant. Most loss across the intake manifold occurs in the valve port's closing duration. The inlet flow area starts decreasing while the mass flow rate follows an increasing trend resulting in a sharp rise in the mean inlet velocity reaching a maximum when the port is about to close fully. The velocity increase causes a corresponding increase in the head loss and pressure drop.

### 4.2 Trapezoidal Port Results

Numerical simulations are performed for the trapezoidal intake manifold under similar boundary conditions as the rectangular port. It is expected that the trapezoidal ports would reduce the pressure losses, especially during closure. The port starts to open from the shorter side. A variation of the pressure drop with rotor angle is illustrated in Figure 12(a-b) for the trapezoidal intake manifold. It is observed that pressure loss is about three times lower than a rectangular geometry. The magnitude of mean inlet velocity during opening is relatively higher in this case compared to the rectangular port, as illustrated in Figure 8(b). Once the port is fully opened, it starts closing from the shorter side. The magnitude of losses is significantly lower during the closing duration of the trapezoidal port as the mean is lower in the trapezoidal port when compared to the rectangular one, as inferred from Figure 8(b).

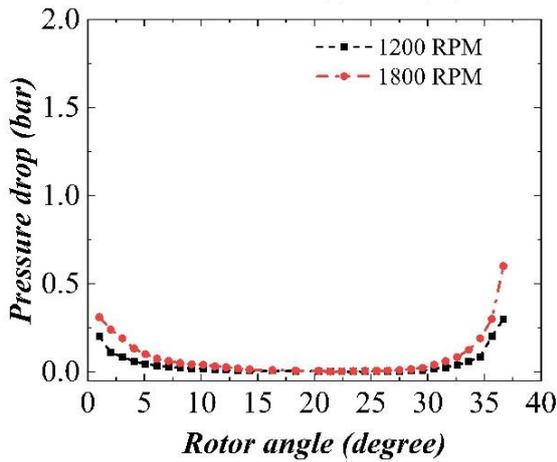 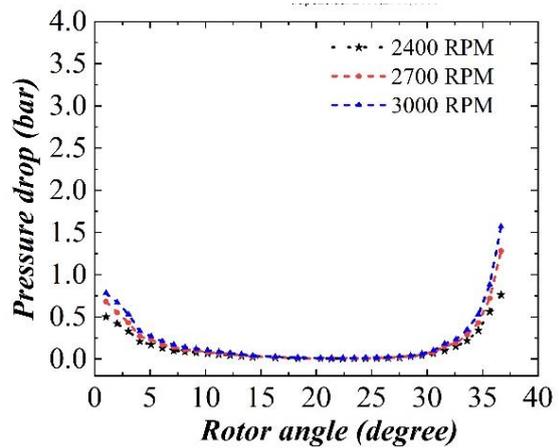

**Figure 12(a):** Pressure drop variation with rotor angle at 1200 and 1800 RPM.

**Figure 12(b):** Pressure drop variation with rotor angle at 2400, 2700 and 3000 RPM.

## 4.3 Pressure Drop and Fluid Flow Analysis during Port Opening and Closing

Figure 13 shows the pressure contours for both profiles for a shaft speed of 3000 RPM at two angular instants during the valve port opening. The decrease in the low-pressure zone at the outlet during the opening is due to the reduction in the mean inlet velocity.

At the onset of port opening (rotor angle of 2°), the area of the low-pressure zone at the outlet of the domain is maximum for both profiles. It starts to fade away gradually with the opening of the port. It is observed that the area of the low-pressure zone at the outlet is higher for the trapezoidal port than the rectangular one, which makes it evident the greater magnitude of losses during the opening. The magnitude of the pressure-drop increases with the shaft speed. A trapezoidal port achieves an average pressure recovery of around 50% during the closing duration.

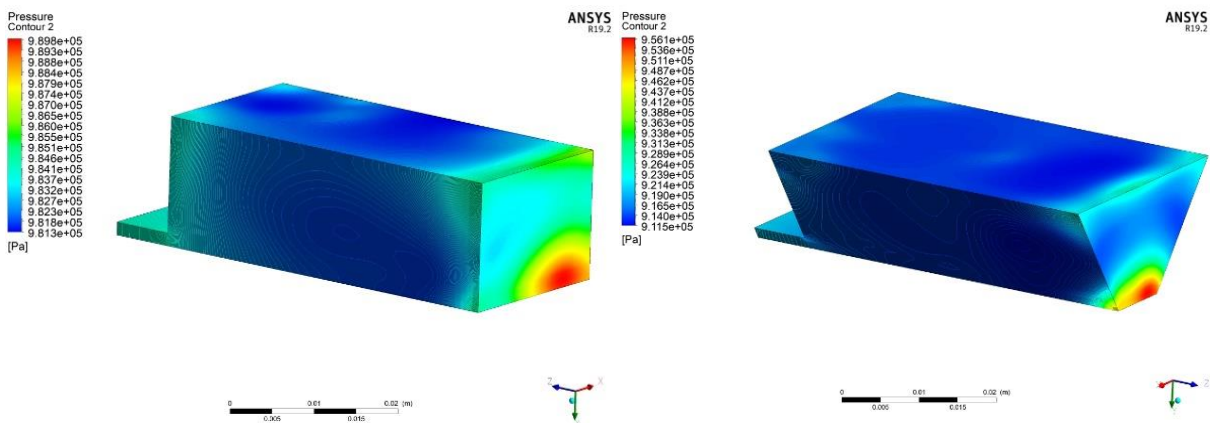

(a)

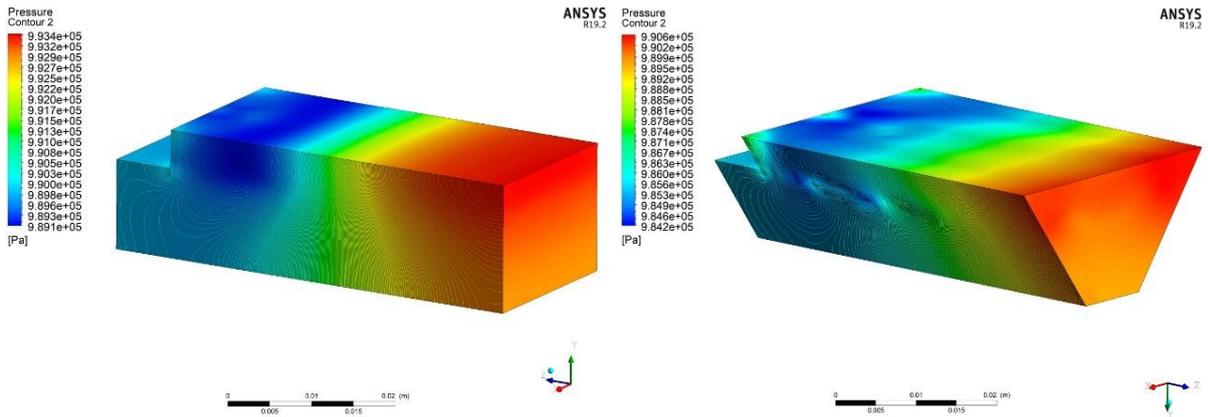

(b)

**Figure 13:** Pressure contours of intake manifold for the rectangular and trapezoidal port at 3000 RPM for a rotor angle of (a): 2° and (b): 14°, respectively.

An illustration of the pressure contours during the port closing duration for both the port configurations is illustrated in Figure 14 for a shaft speed of 3000 RPM. The area of the high-pressure region at the outlet decreases as the port tends to reach a fully closed condition at cut-off. The rectangular manifold encounters higher pressure loss than the trapezoidal manifold. Further, the fluid experiences a larger flow area during closing due to the trapezoid nature, which reduces the magnitude of the wall shear stress for the trapezoidal manifold.

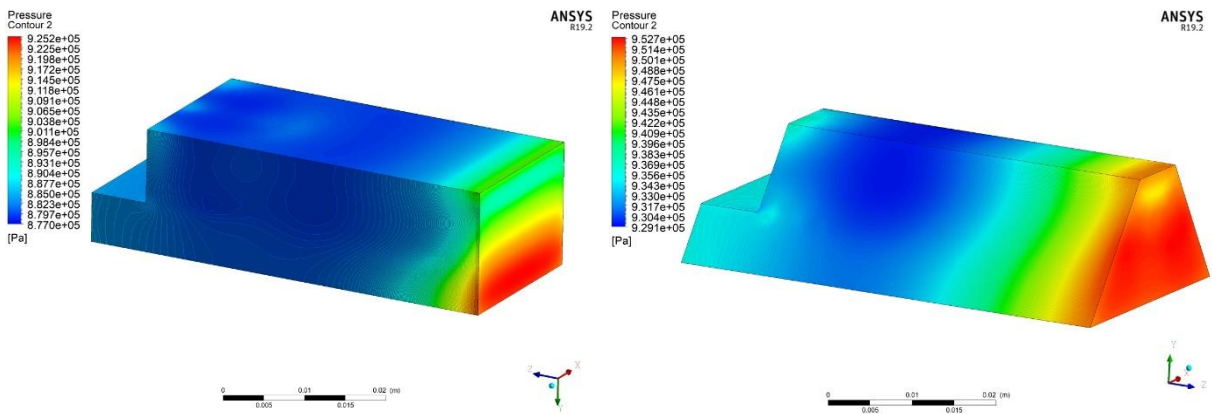

(a)

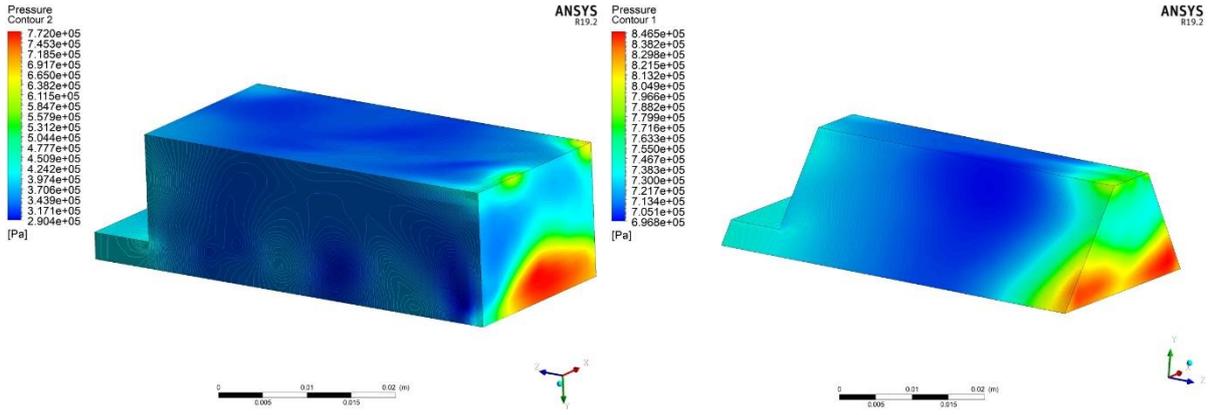

(b)

**Figure 14:** Pressure contours of intake manifold for the rectangular and trapezoidal port at 3000 RPM for a rotor angle of (a): 32° and (b): 36°, respectively.

The flow velocity increases during closing and reaches a maximum before the port fully closes. The high-speed turbulent flow regime results in intense mixing and chaotic motion of the fluid particles across the computational domain. At a shaft speed of 3000 RPM, the portion of the high-pressure region drastically reduces for both the port geometry manifold, resulting in a higher magnitude of pressure drop.

The maximum loss occurs at the valve port's exit during the port's opening and closing duration due to the sudden expansion of the flow domain, which results in high perturbations in the flow field, resulting in an adverse pressure gradient.

### 4.4 Investigation of Wall Shear Stress and Skin Friction Coefficient

The skin friction coefficient, $C_f$ is stated as:

$$C_f = \frac{\tau_{wall}}{\frac{1}{2}\rho V_{m,inlet}^2} \tag{17}$$

The variation of the skin friction coefficient with rotor angle is illustrated in Figure 15 for three different RPMs. The $C_f$ is higher for the trapezoidal manifold compared to the rectangular manifold during the opening duration of the port. As the opening increases, the margin of difference in magnitude of the $C_f$, reduces and converges around midway when the valve port is fully opened during intake.

Variation of mass flow rate during the port opening duration for both the profiles is illustrated in Figure 16 for three RPMs. It follows a decreasing trend during the port opening duration, although the mass flow rate increases continuously as the rate of increase in the flow area exceeds that of mass flux. The shear stress on the wall has a higher magnitude for the trapezoidal manifold than the rectangular manifold during the opening, which agrees with the trend of pressure loss variation for the profile during the opening. As the shaft speed increases, the increase in mean inlet velocity imposes a higher magnitude of the tangential force on the walls of the domain, consequently increasing the shear stress.

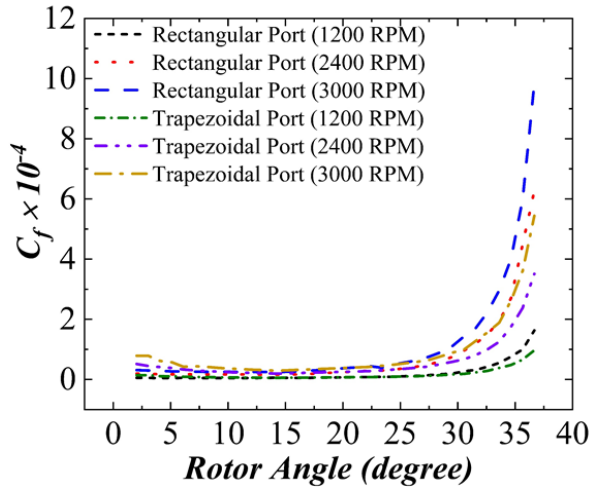 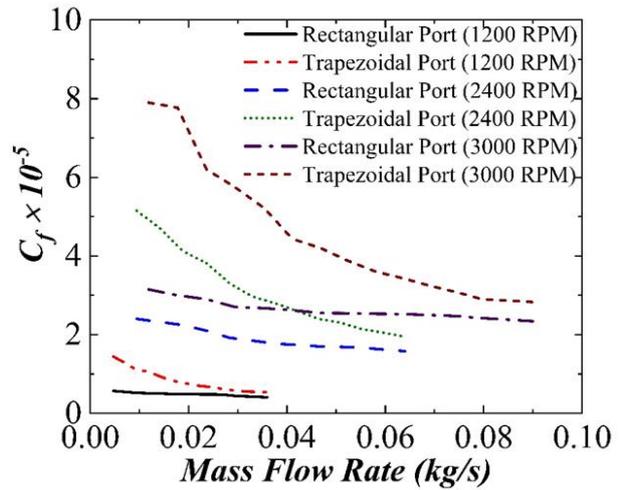

**Figure 15:** Variation of skin friction coefficient with rotor angle at different RPMs.

**Figure 16:** Variation of skin friction coefficient with mass flow rate during the port opening at different RPMs.

Increase in the shear stress and $C_f$ is also observed during port closing due to an increase in mean inlet velocity. A variation of the wall shear stress with the pressure loss during the port closing duration is shown in Figure 17 for a shaft speed range from 1200 to 3000 RPM. The losses increase drastically as the port starts to close, causing the corresponding wall shear stress to increase. Lower pressure losses across trapezoidal manifold during closing reduce wall shear stress's magnitude by 50%. The lesser magnitude of the wall shear for the trapezoidal manifold makes it an efficient port geometry compared to the rectangle.

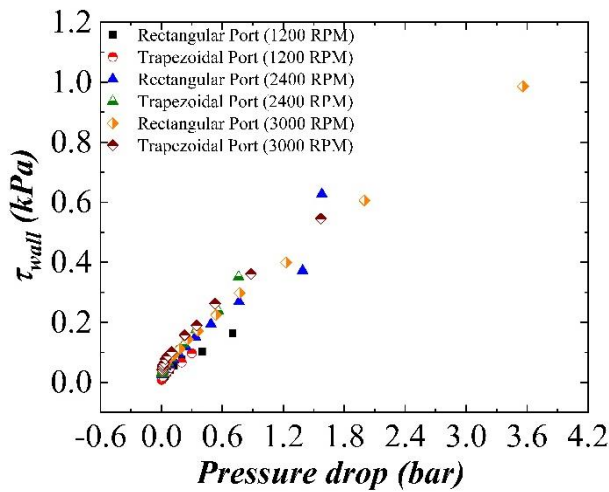 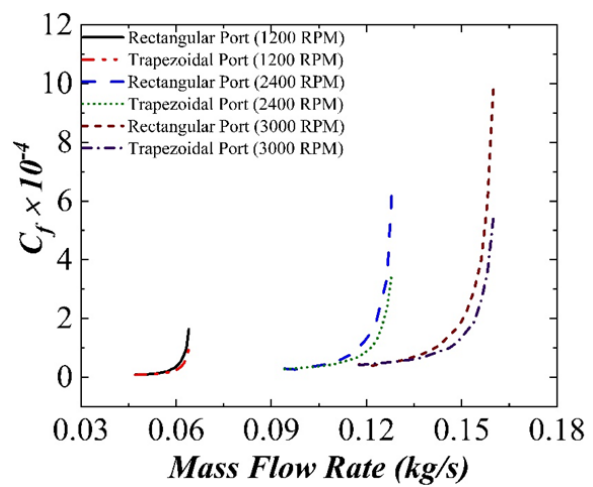

**Figure 17:** Variation of wall shear stress with pressure drop during port closing at different RPMs.

**Figure 18:** Variation of skin friction coefficient with mass flow rate during port closing at different RPMs

We investigate the variation of the $C_f$, with the mass flow rate during the closing duration for the port profile for a range of shaft speed from 1200 to 3000 RPM, as illustrated in Figure 18. The high-speed flow ranging from 27-270 m/s during the closing, in the range of RPMs, imposes a tangential force on the walls of the manifold. The magnitude of $C_f$, follows an increasing trend with a parabolic variation. The higher flow velocity causes the increase in the magnitude of the $C_f$, which increases the pressure drop. The magnitude of the $C_f$ is lower for the trapezoidal manifold than the rectangle for a given shaft speed due to the lower inlet speed for the trapezoidal manifold during the port closing duration.

### 4.5 Turbulence Kinetic Energy (TKE)

The fluid in motion across the intake manifold of the expander has a finite magnitude of kinetic energy. The fluid velocity consists of the mean and fluctuating quantity, respectively. We investigate the kinetic energy associated with these fluctuations, termed turbulence kinetic energy (TKE). As shown in expression (18), a simple time averaging is performed in terms of the flow rate distribution and the mean quantity's magnitude.

$$(V')^2 = \frac{1}{T} \int_0^T (V(t) - \bar{V})^2 \, dt \tag{18}$$

The mean component of the velocity is represented by overbar quantity. The fluctuating quantities in the above expression (18) are subjected to change in space. Figure 19 illustrates the variation of the TKE at the outlet of the intake manifold with rotor angle for a shaft speed of 1200 and 3000 RPM, respectively.

During port opening over a range approximately from 0 to 15 degrees, the TKE decreases for both the ports. However, the magnitude of TKE is higher for the trapezoidal manifold during the opening due to a relatively higher mean inlet velocity magnitude than the rectangular manifold. With the increase in rotor angle, the TKE further decreases and reaches a minimum value approximately at a rotor angle of 15 degrees when the valve port is fully open.

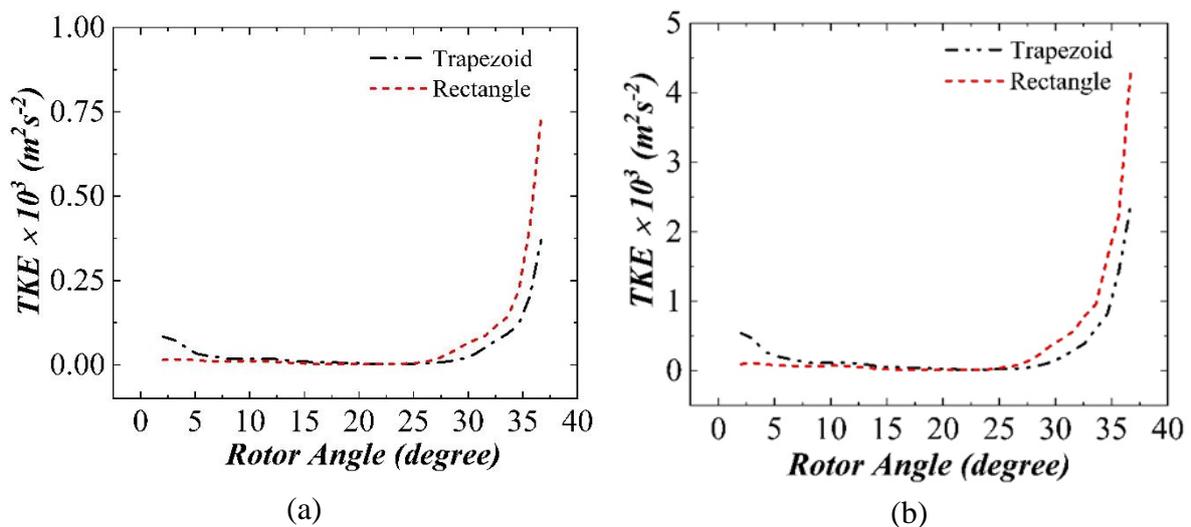

**Figure 19:** Variation of turbulence kinetic energy with rotor angle at (a): 1200 RPM and (b): 3000 RPM, respectively.

It is observed that this angle is independent of the shaft speed of the expander.

A reverse trend is observed during the valve port's closing duration, which ranges from 15 to 36-degree rotor angle. A sharp rise in the TKE is observed during the port closing duration. The inlet flow area decreases as the port starts closing and reaches a minimum value before it gets fully closed. The mass flow rate increases with rotor angle during the entire intake duration, Figure 3, which increases the mean inlet velocity.

### 4.6 Wankel Expander Performance Analysis

#### 4.6.1 Comparison of Net Power Output of the Wankel Expander

Figure 20(a-b) depicts the modified pressure-volume plots for the Wankel Expander, considering the pressure losses across the intake manifold for both the rectangular and trapezoidal port for rotational speeds ranging from 2400 to 3000 RPM. The yellow shaded region illustrates the work potential lost due to the intake losses. For the design pressure ratio, we compare the work potential lost for the expander's ideal work output to investigate the port geometry's standalone effect on net power output. It is observed that the work potential lost due to the intake losses is higher for the rectangular port geometry compared to the trapezoid due to the reduction of the expansion stroke for the rectangular port. The work done in a single cycle is obtained by calculating the enclosed area of the pressure-volume curve using the *polyarea* function (expression 19-22) in MATLAB®. One rotation of the triangular rotor produces three rotations of the eccentric shaft. The net power output of the expander is found using the approach shown below:

$$W_{single\ cycle} = A_{enclosed}(P,V) \tag{19}$$

$$A_{enclosed} = polyarea(P,V) \tag{20}$$

$$P_{out,single\ cycle} = 6 \times (A_{enclosed}) \tag{21}$$

$$P_{net} = P_{out,single\ cycle} \times \frac{N}{3} \tag{22}$$

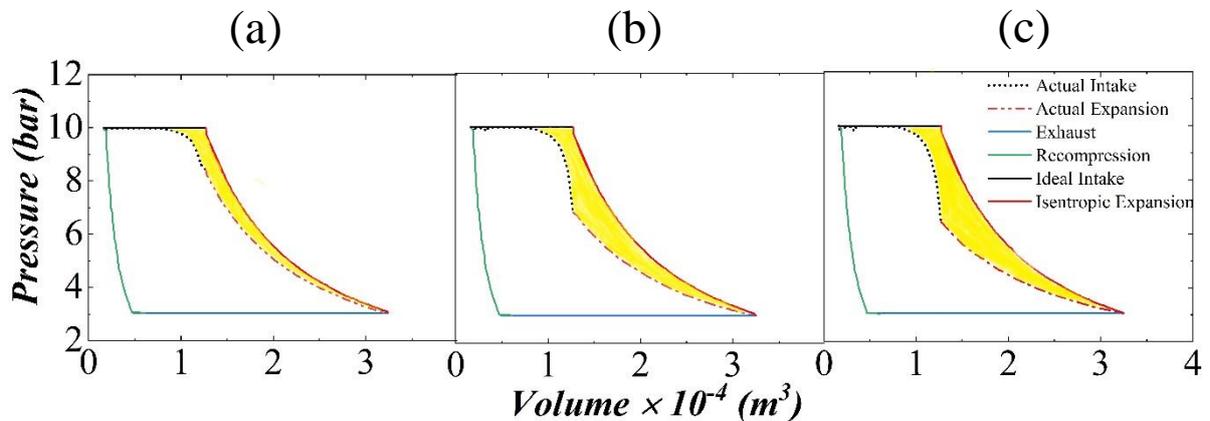

**Figure 20(a):** Non-ideal pressure-volume variation in a single chamber for rectangular port at (a) 2400 RPM, (b) 2700 RPM , and (c) 3000 RPM respectively.

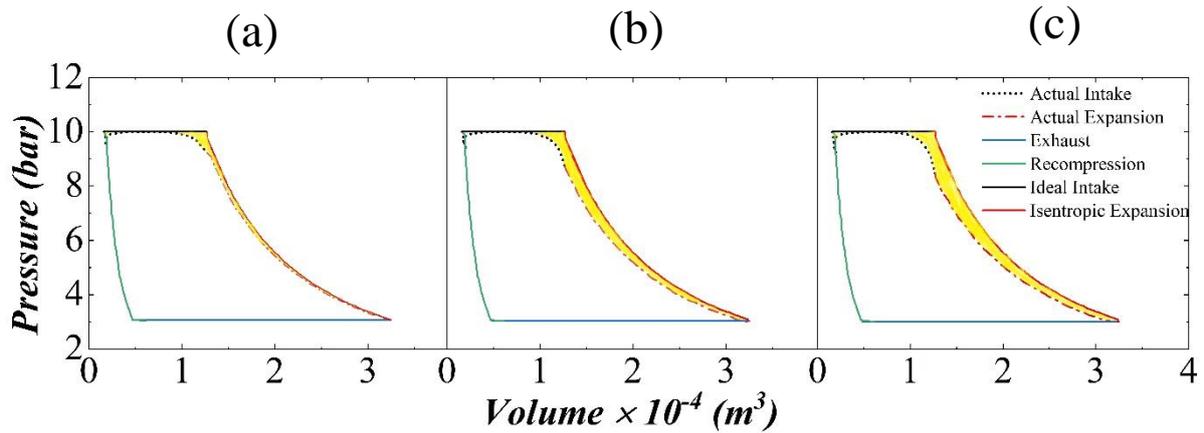

**Figure 20(b)**: Non-ideal pressure-volume variation in a single chamber for trapezoidal port at (a) 2400 RPM, (b) 2700 RPM, and (c) 3000 RPM respectively.

The reduced shaded area in the pressure-volume plots for the trapezoidal profiles in Figure 20(b) indicates less power loss encountered than a rectangular port. Figure 21 illustrates a comparative histogram of the net power output of the expander using the port geometry over a range of RPMs ranging from 1200 to 3000. Higher output power is observed for the trapezoidal port than for the rectangular port. From Figure 22, it is also seen that the percentage increase in power output increases with an increase in shaft speed. Operating the expander at a high shaft speed is necessary to avoid other losses due to steam leakage [19]. An average increment of 15% in the net power output is achieved using the trapezoid geometry in the range of RPMs.

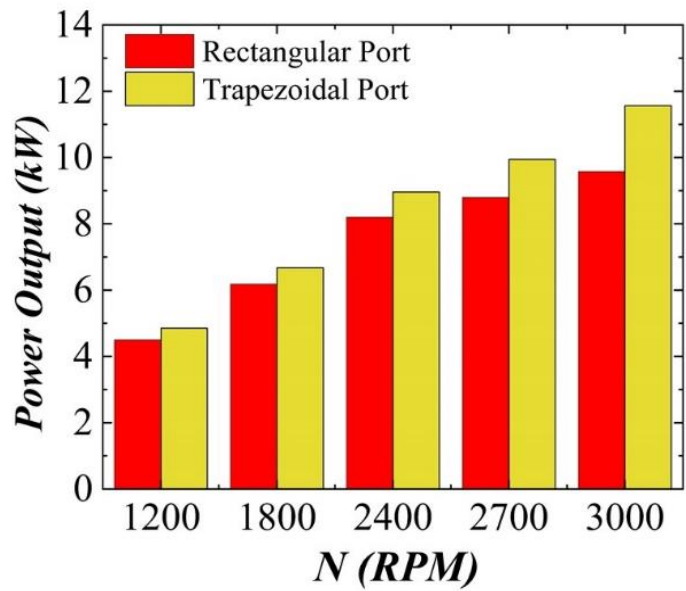

**Figure 21:** Comparative bar chart of power output shaft speed for both port profiles.

### 4.6.2 Effect on Isentropic Efficiency

The theoretical isentropic efficiency of an expander for a given cut-off is assumed to be 100% at the design pressure ratio[12]. We compare the isentropic efficiency considering the losses across the intake manifold for both the port profiles. It is calculated based on the traditional definition of isentropic efficiency of expansion device as stated below[12]:

$$\eta_{is} = \frac{(h_{st,exp} - h_{ed,act\ exp})}{(h_{st,exp} - h_{ed,id\ exp})} \tag{23}$$

Figure 23 illustrates a variation of the isentropic efficiency of the expander with the shaft speed for both the rectangular and trapezoidal intake manifold due to lower intake losses and associated exergy destruction across the trapezoid manifold for a given pressure ratio.

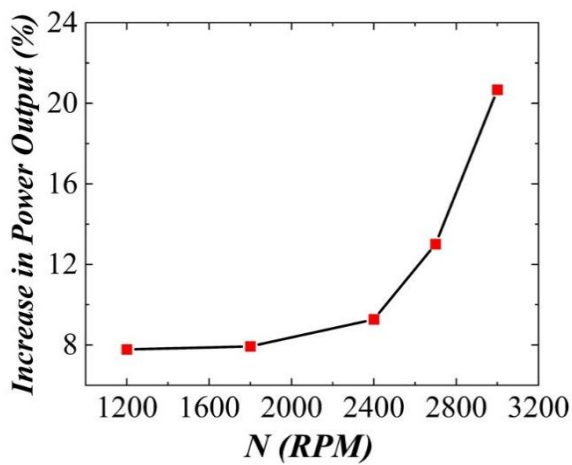 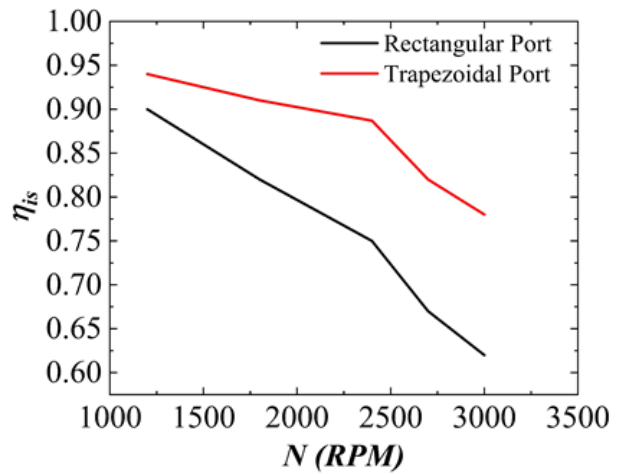

**Figure 22:** Variation of percentage increment in power output with shaft speed using a trapezoidal port.

**Figure 23:** Variation of isentropic efficiency with shaft speed for the port profiles.

## 5. CONCLUSIONS

Numerical simulations of fluid flow across the intake manifold of a Wankel expander were used to estimate the pressure loss for an existing rectangular valve port geometry. Thermodynamic analysis was carried out for the theoretical pressure-volume cycle of the expander and validated with literature. Further, these results were used in a CFD model to estimate the pressure losses across the intake manifold during admission. These intake losses are observed to reduce the power output by 20 to 30% due to changes in the pressure ratio caused by the losses. A trapezoidal port profile of the same hydraulic diameter is designed and proposed for the intake manifold to improve performance. The trapezoidal port reduces the pressure loss during admission by about 50%. Numerical simulations on the performance and power output of the expander were performed to study the effects of the skin friction coefficient, wall shear stress and the turbulence kinetic energy during opening and closing for both the port profiles. An average increment of 15% in the net power output and 14% in isentropic efficiency is achieved using the trapezoid geometry in a shaft speed range from 1200 to 3000 RPM.

**DECLARATION OF COMPETING INTEREST**

The authors declare that they have no known competing financial interests or personal relationships that could have appeared to influence the work reported in the present study.

## ACKNOWLEDGEMENTS

The authors are thankful to the PG Senapathy Computing Resources at IIT Madras for providing the essential computing resources for the present study.

## NOMENCLATURE

| | | |
|---|---|---|
| $A$ | Area | -- |
| $b$ | Rotor Width | mm |
| $e$ | Eccentricity | mm |
| $f$ | Friction factor | |
| $h$ | Enthalpy | J/kg |
| $k$ | Turbulence Kinetic Energy | $m^2/s^2$ |
| $N$ | Shaft Speed | RPM |
| $O$ | Origin | -- |
| $P$ | Pressure | $N/m^2$ |
| $R$ | Rotor radius | mm |
| $t$ | Time | s |
| $\bar{V}$ | Mean Velocity | m/s |
| $V$ | Volume | $mm^3$ |
| $W$ | Work | J |
| $x$ | Coordinates along the x-axis | |
| $y$ | Coordinates along the y axis | -- |
| $z$ | Coordinates along the z-axis | |

*Greek Symbols*

| | | |
|---|---|---|
| $\theta$ | Shaft Angle (°) | |
| $\upsilon$ | Rotor Angle (°) | |
| $\eta$ | Efficiency (%) | |
| $\varepsilon$ | dissipation rate ($m^2/s^3$) | |
| $\rho$ | Density ($kg/m^3$) | |
| $\mu$ | Dynamic Viscosity ($Ns/m^2$) | |
| $\Delta$ | Surface roughness factor | |

*Subscript/Superscript*

| | |
|---|---|
| $b$ | boiler |
| $c$ | condenser |
| $m$ | mean |
| $ad$ | admission |
| $in$ | inlet |
| $net$ | Net power output |
| *st,exp* | Start of the expansion process |
| *ed, act exp* | End of the actual expansion process |
| *ed, id exp* | End of the ideal expansion process |